\documentclass[preprint,12pt]{elsarticle}

\usepackage{amsmath}
\usepackage{amssymb}
\usepackage{bm}
\usepackage{graphicx}
\usepackage{booktabs}
\usepackage{hyperref}

\journal{Acta Materialia}

\biboptions{sort&compress}

\DeclareUnicodeCharacter{E5F8}{}

\begin{document}
\begin{frontmatter}

\title{Origin of the anomalous shape of the $\gamma$-loop in paramagnetic Fe$_{1-x}$Cr$_x$ alloys\\ from coupled Atomistic Spin Dynamics and \emph{ab initio} Molecular Dynamics}

\author[ifm,wise]{Joakim Stenborg\corref{cor1}}
\ead{joakim.stenborg@liu.se}

\author[ifm,helsinki]{Davide Gambino}

\author[ifm,wise]{Bj\"{o}rn Alling}

\cortext[cor1]{Corresponding author}

\affiliation[ifm]{
    organization={Department of Physics, Chemistry and Biology (IFM),
                  Link\"{o}ping University},
    postcode={SE-581 83},
    city={Link\"{o}ping},
    country={Sweden}
}

\affiliation[wise]{
    organization={Wallenberg Initiative Materials Science
                  for Sustainability (WISE),
                  Link\"{o}ping University},
    city={Link\"{o}ping},
    country={Sweden}
}

\affiliation[helsinki]{
    organization={Department of Physics,
                  University of Helsinki},
    addressline={P.O. Box 43},
    postcode={FI-00014},
    city={Helsinki},
    country={Finland}
}

\begin{abstract}
The region of fcc stability in Fe-rich Fe-Cr alloys, the $\gamma$-loop, has an anomalous shape: at low temperatures, Cr addition initially stabilizes fcc, but above 8 at.\% Cr it stabilizes bcc. At high temperatures, Cr always stabilizes bcc. This behavior reflects a delicate balance between magnetic, vibrational and configurational free-energy contributions that is challenging to capture from first principles. We investigate phase stability in Fe$_{1-x}$Cr$_x$ ($x=0, 0.05, 0.10, 0.15$) using coupled atomistic spin dynamics and $ab$ $initio$ molecular dynamics with thermodynamic integration. Gibbs free energy differences between fcc and bcc are calculated from 800 to 1800 K. Our calculations reproduce the experimental $\gamma$-loop and its compositional dependence, with near-quantitative accuracy for the $\gamma-\delta$ transition. The $\alpha-\gamma$ boundary, including its anomalous shape, is reproduced qualitatively, although transition temperatures are underestimated by approximately 200 K. Static calculations of magnetically ordered and disordered states provide an explanation: in ordered states, increasing Cr initially reduces the energetic preference for bcc over fcc, whereas in disordered states, Cr progressively stabilizes bcc at all compositions. Vibrational effects contribute to phase stability, while atomic short-range clustering at higher Cr content ($x\geq 0.15$) may explain the observed closing of the $\gamma$-loop from below.

The simulations reveal finite Cr moments in the paramagnetic state, whose magnitudes depend on their local magnetic environment. The results demonstrate that the anomalous $\gamma$-loop arises from the interplay of magnetic energetics, magnetic disorder, lattice vibrations and chemical clustering, which must be treated on equal footing to capture phase stability from first principles.
\end{abstract}


\end{frontmatter}

\section{Introduction}
Predicting the stability of materials under varying thermodynamic conditions is a foundational challenge in materials science \cite{dehoff2006thermodynamics}. A detailed understanding of why one particular atomic arrangement is favored over another under specific temperature, pressure, and compositional regimes is essential for the design of structural alloys \cite{porter2009phase}. Achieving this goal requires computational methods that can accurately capture the effect on free energy of the complex interplay of crystall structure, chemical composition, atomic configuration, and electronic and magnetic degrees of freedom, particularly at elevated temperatures \cite{de1994cluster, van2002automating, Ruban_Abrikosov_alloy_review_2008, fultz2010vibrational, glensk2015understanding, Abrikosov_PM_magnetism_review_2016, kormann2016influence}.  Magnetic alloys represent a challenging class of materials in this context \cite{staunton1984disordered, gyorffy1985first, alling2006role, alling2010theory, kormann2014lambda, ehn2023first}.  Above, and particularly just above, the magnetic ordering temperature these systems can exhibit strong coupling between atomic vibrations, chemical disorder, and fluctuating magnetic moments \cite{mozafari2015role,  Stockhem2018Anomalous, hickel2025magnetism, biswas_decoding_2026}. These intertwined effects give rise to rich phase behavior that is highly sensitive to temperature and composition \cite{lavrentiev2010magnetic, kormann2011role, ruban2012spin}. Even for pure Fe, despite beeing of the highest technological importance for thousands of years, a fully $ab$ $initio$ reproduction of the metal's $\alpha-\gamma-\delta$ transition has just recently been demonstrated.~\cite{Gambino2023Phase} 

A prime example of the complexity of magnetic alloys is the iron-chromium (Fe$_{1-x}$Cr$_x$) system, which forms the basis of stainless steels \cite{lula1985stainless} and serves as a key structural material in nuclear reactor components \cite{dudarev2009eu}. The Fe-Cr phase diagram exhibits a variety of structural and magnetic phase transitions \cite{xiong2010phase, jacob2018revised}, with one of the most intriguing features being the anomalous shape of the so-called $\gamma$-loop. The $\gamma$-loop is a general feature of Fe alloyed with body-centered cubic (bcc)-favouring elements \cite{predel1991phase}, and consists of a region in which the face-centered cubic (fcc, $\gamma$) phase is thermodynamically stable only within a restricted composition and temperature range, entirely enclosed by a region of bcc ($\alpha$ and $\delta$) stability \cite{jacob2018revised}. Typically, increasing the concentration of the alloying element suppresses the stability of the fcc phase at both high and low temperatures. 
In Fe-Cr, however, the expected bcc stabilization with increasing Cr content is observed only for high temperatures, at the $\gamma-\delta$ transitions. For low temperatures, at the $\alpha-\gamma$ transition, the fcc phase is instead progressively stabilized up to approximately 8 at.\% Cr, beyond which the more conventional bcc stabilization resumes. 
Despite several theoretical studies of Fe$_{1-x}$Cr$_x$ system  \cite{olsson2003ab, mirebeau2019magnetic, korzhavyi2009electronic, lavrentiev2012magnetic, yang2024insight, casillas2023effect}, the fundamental origin of this anomalous behavior of the $\gamma$-loop is not completely understood, and computational reproduction of the fcc stabilization at the $\alpha-\gamma$ transition has not yet been reported.

Despite the tremendous improvements in first-principles thermodynamic calculations occurred in recent years \cite{grabowski2009abinitio,glensk2014breakdown,glensk2015understanding}, the fact that the bcc-fcc-bcc transition in this system occurs entirely within the paramagnetic state poses a great challenge. Traditional first-principles methods, including density functional theory (DFT), often treat magnetic and structural degrees of freedom separately \cite{ruban2012first, drautz2004spin}. As a result, they struggle to predict finite-temperature phase stability in magnetic alloys, especially in disordered or high-temperature regimes.
In pure Fe, this high-temperature state is known to display a delicate balance between magnetic and vibrational contributions to the free energy, \cite{Gambino2023Phase, pepperhoff2013constitution, kormann2014temperature}, and the complexity increases further with composition-dependence and configurational effects with Cr addition. Solution to the puzzle of the $\gamma$-loop in Fe$_{1-x}$Cr$_x$ requires accurate methods that can treat simultaneosly all degrees of freedom and resolve free energy differences in the order of a few meV. 

In this work, we use a first-principles-based  methodology that explicitly incorporates \emph{spin-lattice coupling}, the mutual interaction between magnetic and atomic degrees of freedom, into finite-temperature thermodynamic calculations. By employing  combined Atomic Spin Dynamics (ASD) - \emph{Ab Initio} Molecular Dynamics (AIMD) simulations, as developed in Ref.~\cite{Stockhem2018Anomalous}, we access the electronic, magnetic, and vibrational state of the alloys. Then, by thermodynamic integration over temperature, volume and stress-strain variables, we compute the Gibbs free energy difference between bcc and fcc Fe$_{1-x}$Cr$_x$ alloys across a range of chromium concentrations. This approach has previously been shown to accurately capture the $\alpha-\gamma-\delta$ transition in pure Fe \cite{Gambino2023Phase}, and allows us to systematically account for the effects of magnetic disorder and atomic vibrations, yielding new insights into the mechanisms that stabilize different phases in the Fe$_{1-x}$Cr$_x$ system. Indeed, our first-principles based simulations reproduce the $\gamma$-loop anomaly, enabling a detailed investigation of its microscopic origin. Our results demonstrate that predictive, first-principles based thermodynamic modeling of steels and other complex magnetic materials under synthesis- and application-relevant conditions is now within reach.

\section{Method}

In the present study, DFT calculations were performed using the Vienna Ab initio Simulation Package (VASP) \cite{kresse1996efficient}, while Monte Carlo (MC) and Atomistic Spin Dynamics (ASD) simulations were carried out with the Uppsala atomistic spin dynamics package (UppASD) \cite{eriksson2017atomistic}. VASP calculations employed the projector-augmented wave (PAW) method \cite{PAW1,PAW2} and plane waves as the basis set, and the Perdew-Burke-Ernzerhof (PBE) \cite{PhysRevLett.77.3865} generalized gradient approximation for the exchange-correlation functional. A fully noncollinear description of the magnetization density was employed.

All AIMD simulations were performed with a timestep of 1 fs for a total duration of 5-10 ps, in an NVT ensemble using a Langevin thermostat with a friction coefficient of 10 ps$^{-1}$. Each simulation employed a 125-atom supercell, with plane-wave cutoff of 250 eV and $\Gamma$-point sampling. Chemical disorder was modelled in these supercells with a special quasirandom structure (SQS) \cite{zunger1990special}. The 125 supercells for fcc and bcc were represented in a body-centered tetragonal (bct) setting, which allows for a continuous transformation from bcc to fcc along the Bain path \cite{bain1924nature}. At intervals of approximately 200 fs, snapshots were recalculated (``upsampled") using a higher plane-wave cutoff of 500 eV and a $2 \times 2 \times 2$ k-points mesh. For each case, the average energy and stress tensor from the lower-accuracy simulations were corrected with the mean difference between the high- and low-accuracy results in the spirit of thermodynamic perturbation theory, as suggested in Ref. \cite{Upsampling.method.Dorner}. The validity of the upsampling scheme was verified for pure Fe in Ref. \cite{Gambino2023Phase}. 

The magnetic moments in the system were interpreted as noncollinear three-dimensional vectors located at each atomic site, obtained by integrating the magnetization density vector field up to radii of 1.397 Å for Fe and 1.323 Å for Cr around each atom. For all DFT simulations, we employed the constraining method of Ma and Dudarev \cite{ma2015constrained} to constrain the direction and sign, but not the magnitude, of Fe moments, while the moments of the Cr atoms were initialized to zero and allowed to evolve during Kohn Sham-loops according to the local environment. This approach was chosen due to the low Cr concentration and the dominant role of the Fe moments on the Cr moments, as well as to avoid convergence issues associated with constraining the weak Cr moments. Longitudinal spin fluctuations of Fe moments are included in the VASP simulations adding an entropic contribution of the type $S_{mag}=k_B \log(m)$ to the single-particle wavefunctions~\cite{ruban2013impact}. The effects of adding this entropic contribution to the Cr atoms as well was also investigated for Fe$_{85}$Cr$_{15}$ (See the Supplementary Material).

Between each AIMD timestep, the Fe moments were updated according to the Landau-Lifshitz-Gilbert (LLG) equations \cite{landau1935theory, gilbert2004phenomenological} within the ASD simulations. The coupling of AIMD with ASD, referred to hereafter as ASD–AIMD simulations, enables the simultaneous treatment of lattice vibrations and spin dynamics. The ASD simulations were performed in smaller timesteps of $10^{-2}$ fs, while spanning the same total duration of 1 fs as the corresponding AIMD step. Only the Fe moments were updated during this procedure, using distance-dependent Fe-Fe magnetic exchange interactions from Ref.~\cite{Gambino2023Phase} for fcc and Ref.~\cite{Yin.exchange.interactions} for bcc, ensuring proper coupling with the MD simulation. A test of the effect of including Fe–Cr magnetic exchange interactions in bcc was also made in the Fe$_{85}$Cr$_{15}$ simulations, the details of which can be found in the Supplementary Material; however, their effect on the Gibbs free energy difference was found to be minor and was therefore not pursued further.

To evaluate the Gibbs free energy difference between fcc and bcc, three types of thermodynamic integration were employed: stress–strain thermodynamic integration (SSTI), temperature thermodynamic integration (TTI), and volume thermodynamic integration (VTI). A detailed description of the corresponding formulas and procedures can be found in Ref.~\cite{Gambino2023Phase}. SSTI consists in the integration of thermodynamic stresses along the deformation path between two lattices \cite{klarbring2018phase}. For all compositions considered in this study, five points along the fcc–bcc Bain path were used: the fcc and bcc structures as endpoints, together with three intermediate structures generated by linear interpolation of the cell along the deformation path. All five structures were simulated with disordered local moment (DLM) \cite{gyorffy1985first}-AIMD simulations \cite{alling2016strong} at 1400~K, far above the magnetic ordering temperature but well below the melting point. 
This temperature ensures consistent treatment of the magnetic degrees of freedom between the different structures and prevents excessive diffusion in the intermediate geometries, which would require simulations over much longer timescales to obtain well-converged quantities. 
DLM-AIMD simulations were performed in the same manner as the ASD-AIMD simulations, except that in the ASD step the exchange interactions were set to zero and the temperature was fixed at $10^{6}$ K, which effectively re-randomised the moments at each timestep.

The resulting fcc–bcc free energy differences were then extended up to 1800~K using TTI, following the formula presented in Ref.~\cite{cheng2018computing}. For this purpose, bcc and fcc DLM–AIMD simulations were then performed at 1600~K and 1800~K at the same volume as at 1400~K, ensuring that temperature was the only varying contribution to the free energy difference. At 1800~K, thermal disorder of the magnetic state is sufficiently large that ASD–AIMD and DLM–AIMD yield equivalent results, allowing the free energy difference to be transferred from DLM–AIMD to ASD–AIMD. ASD–AIMD simulations for both bcc and fcc were performed at 1800~K, 1400~K, 1000~K, and 800~K for the downward TTI. To determine the Gibbs free energy, additional VTI simulations at varying volumes were performed at each temperature in order to find the equilibrium volume and the free energy contribution from volume difference.

In addition, to gain clarity in the energetic origin of the observed effects, static VASP SQS calculations of 60 atoms supercells at x=0, 0.05, 0.10, 0.15, 0.20, 0.25, 0.30 and 0.50 was done for both fcc and bcc structures. FM and DLM magnetic states were considered, as well as the noncollinear magnetic ground-state for fcc structures at x=0.00, 0.10 and 0.20 obtained using the method proposed by Ehn $et$ $al.$ \cite{ehn2023first}. In these static calculations the equilibrium volumes obtained from ASD-AIMD at 1000 K was used, or, in the case of Cr-rich compositions, extrapolations from the ASD-AIMD considered compositions was done. In the DLM calculations the FM-relaxed atomic positions were used.

To check the effect of possible short-range clustering (SRC) in the bcc structure, we perform additional Monte-Carlo (MC) simulations following the methods described in Ref.~\cite{ruban2012first, korzhavyi2009electronic}. Using the first three dominant  pair-interactions $V_{ij}$ in the DLM state derived there for x=0.15, we perform atomic MC simulations and obtain configurational free energies from 1800 K down to 700 K. From these, we obtain temperature-dependent short-range-order parameters, configurational energies, and, through thermodynamic integration from the high-temperature limit, temperature-dependent entropies of mixing. 

We compare the resulting short-range-order parameters, which indicate a tendency toward clustering, with the values experimentally obtained in Ref.~\cite{mirebeau2019magnetic}. The experimental values are slightly larger compared to those obtained from the MC simulations using the theoretical $V_{ij}$.

It should be noted that in the theoretical simulations, the dominant clustering effect occurs in the first coordination shell, while the second shell exhibits significantly weaker clustering. This distinction is difficult to resolve experimentally when averaging over the first two shells, likely due to the small radial distance difference between them in the bcc structure.

The experimental $\alpha_{12}$ values can instead be reproduced using MC simulations with the interaction parameters from~\cite{ruban2012first, korzhavyi2009electronic} scaled by a factor of $1.1$ at $T = 1100\,\mathrm{K}$ and $1.3$ at $T = 700\,\mathrm{K}$. To estimate the effect of experimentally observed SRC on the configurational free energies, we compute the free energies using MC simulations with these scaled interactions and interpolate between them.

For $700\,\mathrm{K} < T < 1100\,\mathrm{K}$, the configurational free energy is given by:
\begin{equation}
F^{conf} = (1 - y) F_{1.3}^{conf} + y F_{1.1}^{conf}, \quad y = \frac{T - 700}{400}.
\end{equation}
For $T > 1100\,\mathrm{K}$,
\begin{equation}
F^{conf} = F_{1.1}^{conf}.
\end{equation}

\section{Results and discussion}
\subsection{Thermal Expansion}

\begin{figure}
    \centering
    \includegraphics[width=0.96\textwidth]{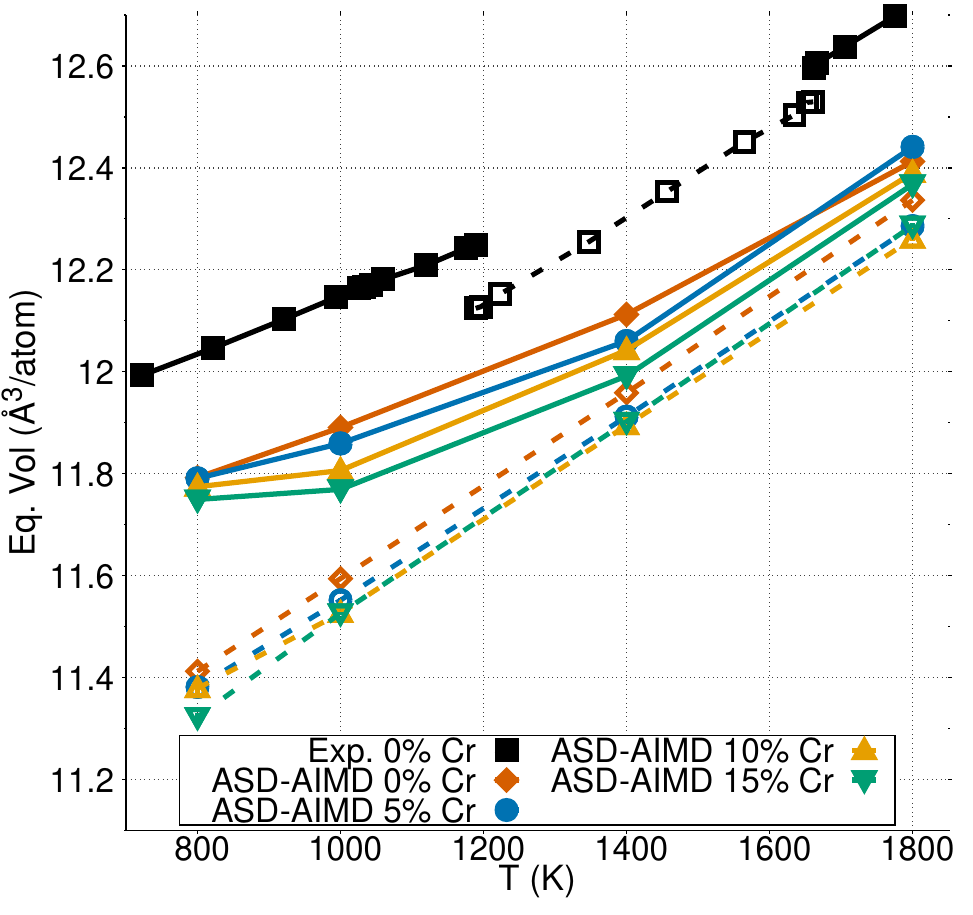}
    \caption{Calculated ASD-AIMD equilibrium volumes of fcc (dashed lines) and bcc (solid lines) phases as a function of temperature for each composition. Experimental data \cite{basinski1955lattice} are shown for pure Fe.}
    \label{fig:lat_v_T}
\end{figure}

The temperature-dependent equilibrium volume was derived from the ASD-AIMD simulations for each composition and both fcc and bcc structures. In each case, using the sampling of different volumes in the VTI-step of our work-flow, the volume corresponding to zero pressure was obtained. These resulting equilibrium volumes are plotted as a function of temperature in Fig.~\ref{fig:lat_v_T}. The experimental volumes for pure Fe are shown for comparison.

In fcc, the temperature change of the equilibrium volumes is practically linear. For pure Fe, comparison with experiments for $\gamma$-Fe shows an under-estimation in our calculations of about 3\%, independent of temperature. The addition of Cr tends to reduce the volume, especially between pure Fe and Fe$_{95}$Cr$_{05}$.

In the bcc phase, experiments show that thermal expansion is lower in the $\alpha$-phase compared to the $\delta$-phase. Our calculations are in line with this finding, although we do not have a detailed resolution of temperatures above the $\gamma-\delta$ transition. With increasing Cr-content, we observe a larger difference in expansion between low temperatures, around 800-1000 K, and high temperatures, between 1400-1800 K. At low T, the expansion gradually decreases with Cr content, while at high T, the expansion instead is higher, making the equilibrium volumes again less sensitive to composition. We will return to the origin of this effect in the discussion of magnetic moments below.

A qualitatively similar composition- and temperature-dependent change in thermal expansion has been reported in the experimental study of $\alpha-$ and $\gamma-$Fe$_{1-x}$Cr$_x$ by Babyuk et al.  \cite{etde_5088552}.  They observed that Fe$_{1-x}$Cr$_x$ alloys with 6 and 8 at.\% Cr showed a distinct decrease in thermal expansion near their Curie temperatures. Their measurements also demonstrated that thermal expansion was higher in Cr-poor compositions than Cr-richer compositions up to around 1000~$K$, but that the slope of the curves indicates that Cr-richer compositions would overtake Cr-poor compositions in thermal expansion at higher temperature, although the $\delta$-phase was not studied explicitly. Qualitatively, these findings are in line with our theoretical results, but quantitatively, our simulations seem to exaggerate the amplitude of the non-linearity of thermal expansion in bcc Fe$_{1-x}$Cr$_x$ at higher x. We believe that this is partly due to our approximation of only including the LSF term on Fe atoms. We test this hypothesis by also including the LSF term in the Cr moments; indeed, a slight reduction in the amplitude of the non-linear effect of thermal expansion is observed (see Supplementary Material).  

\subsection{Magnetic structure}

In all ASD-AIMD and DLM-AIMD simulations, finite Cr moments were found. Previous studies investigating paramagnetic Fe-Cr with static-lattice DLM have found vanishing Cr moments \cite{al2011bain}. Ref.~\cite{al2011bain} used the coherent potential approximation (CPA) for the magnetic disorder in the system, which neglects local environment effects due to its single-site nature. Basically, in DLM-CPA calculations, every magnetic moment sees an identical cancellation of the net spin direction of all local environments. In our supercell approach, on the other hand, we explicitly include such effects, thereby enabling the emergence of a finite Cr local magnetic moment as a consequence of its specific atomic and magnetic environment within the supercell at each snapshot in time. It is our view that this is indeed a better approximation of the real paramagnetic state than the mean field single site approximation \cite{khan2017local, yoon2019non, alling2010effect}. To our knowledge, no other studies, either experimental or computational, have reported the presence of Cr moments in the high-temperature paramagnetic state of Fe-Cr.

\begin{figure}[]
    \centering
    \includegraphics[width=0.90\textwidth]{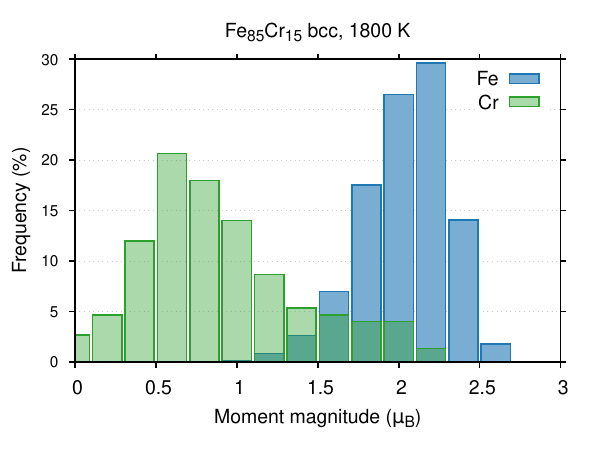}
    \includegraphics[width=0.90\textwidth]{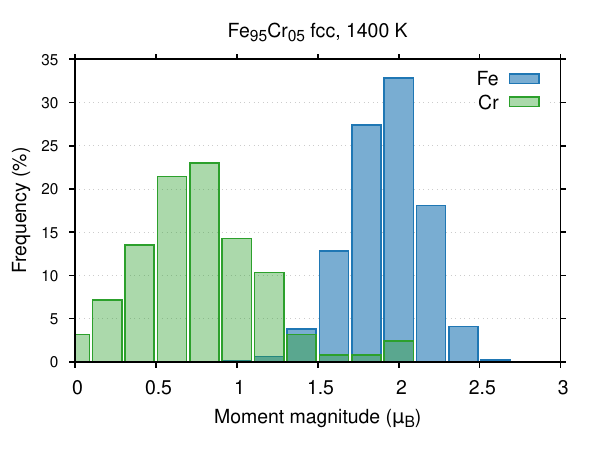}
    \caption{Distribution of moment magnitude for Fe$_{95}$Cr$_{05}$ at 1800K in the bcc lattice and 1400K in the fcc lattice.}
    \label{fig:Mom_hist}
\end{figure}

Fig.~\ref{fig:Mom_hist} shows the distribution of moment magnitudes from Fe$_{95}$Cr$_{05}$ in fcc at 1400 K and bcc at 1800 K simulations at approximate equilibrium volume. These conditions were selected because the respective crystal structures are thermodynamically stable at these temperatures, according to both the experimental phase diagram and our simulations. The resulting distributions may therefore provide insight into the magnetic behavior of Fe–Cr alloys under experimentally relevant conditions. In all simulations, we observe substantial variation in the Cr moments, whereas the Fe moments remain comparatively stable. Since both cases in Fig.~\ref {fig:Mom_hist} are highly paramagnetic, as enforced by the constraints on the Fe atoms, the persistence of well-defined Fe moments suggests that the Fe local moments are energetically robust, corresponding to a deep minimum in the magnetic energy landscape. By contrast, the much broader distribution of Cr moments indicates that the Cr moment magnitude is less well defined and more sensitive to the local environment, consistent with a flatter magnetic-energy landscape for Cr.

\begin{figure}
    \centering
    \includegraphics[width=0.70\textwidth]{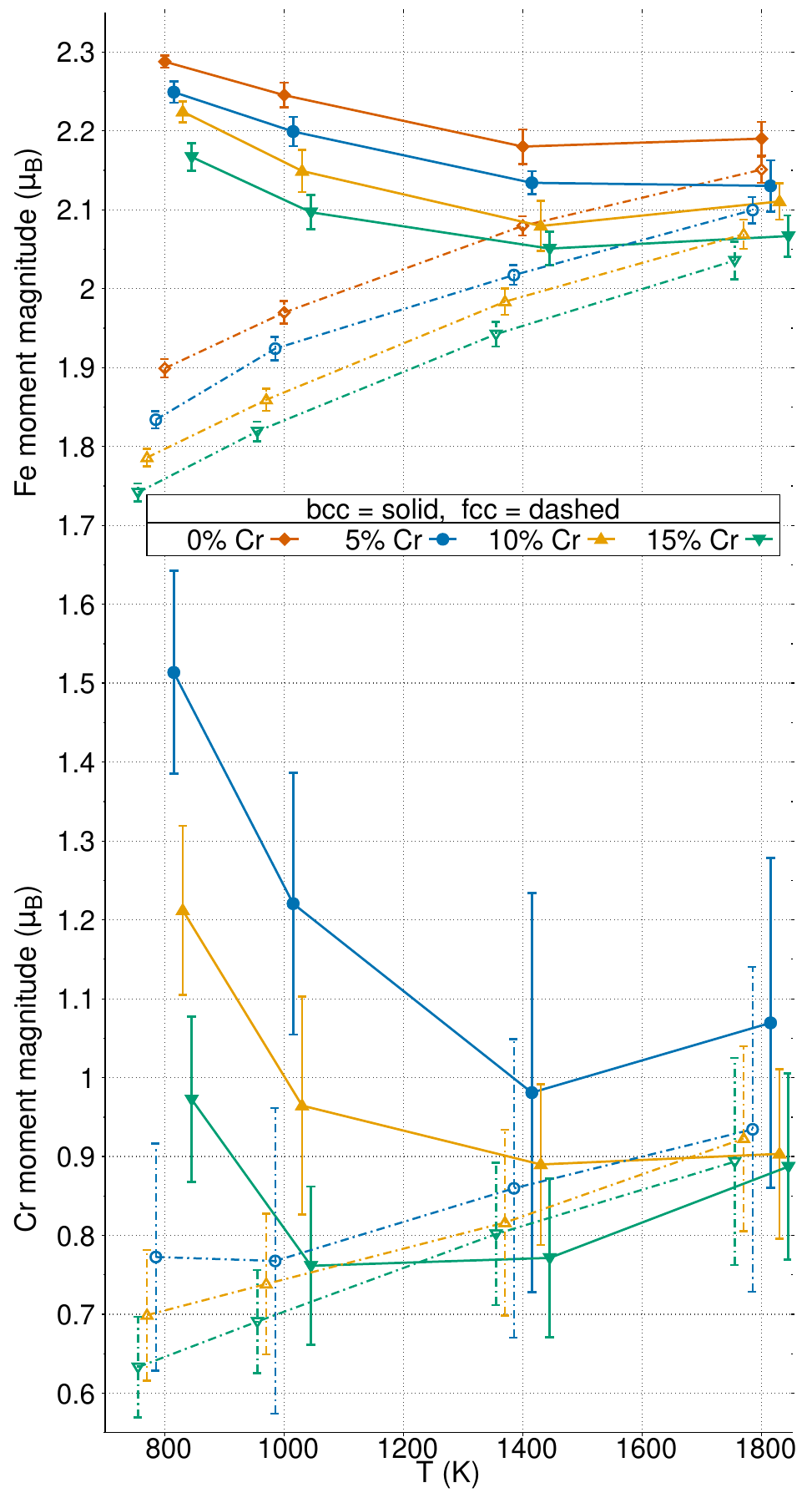}
    \caption{Fe and Cr average moment magnitude versus temperature for all compositions and for fcc (dashed) and bcc (solid). Temperatures for 5\%, 10\%, and 15\% Cr in bcc (fcc) have been shifted by 15, 30, and 45 K to the right (left), respectively, for easier reading.}
    \label{fig:mom_mag}
\end{figure}

Fig.~\ref{fig:mom_mag} shows the average magnetic moment magnitude of Fe and Cr at zero pressure as a function of temperature, composition, and lattice type. The values were either obtained directly from the simulations at equilibrium lattice parameter, or determined by interpolating or extrapolating the calculated moments from simulations at different lattice parameters. For values obtained directly from simulations, the error bars correspond to the standard deviation of the local moment distribution. For values obtained by interpolation or extrapolation, the uncertainty was estimated using an envelope-based method, in which upper and lower bounding curves were constructed from the error bars of the simulated data points used for interpolation or extrapolation. The resulting spread was used to define the error bars of the interpolated or extrapolated values. Large error bars are found for Cr moments due to the comparatively small amount and the large variation in their size. 

\begin{figure}
    \centering
    \includegraphics[width=0.96\textwidth]{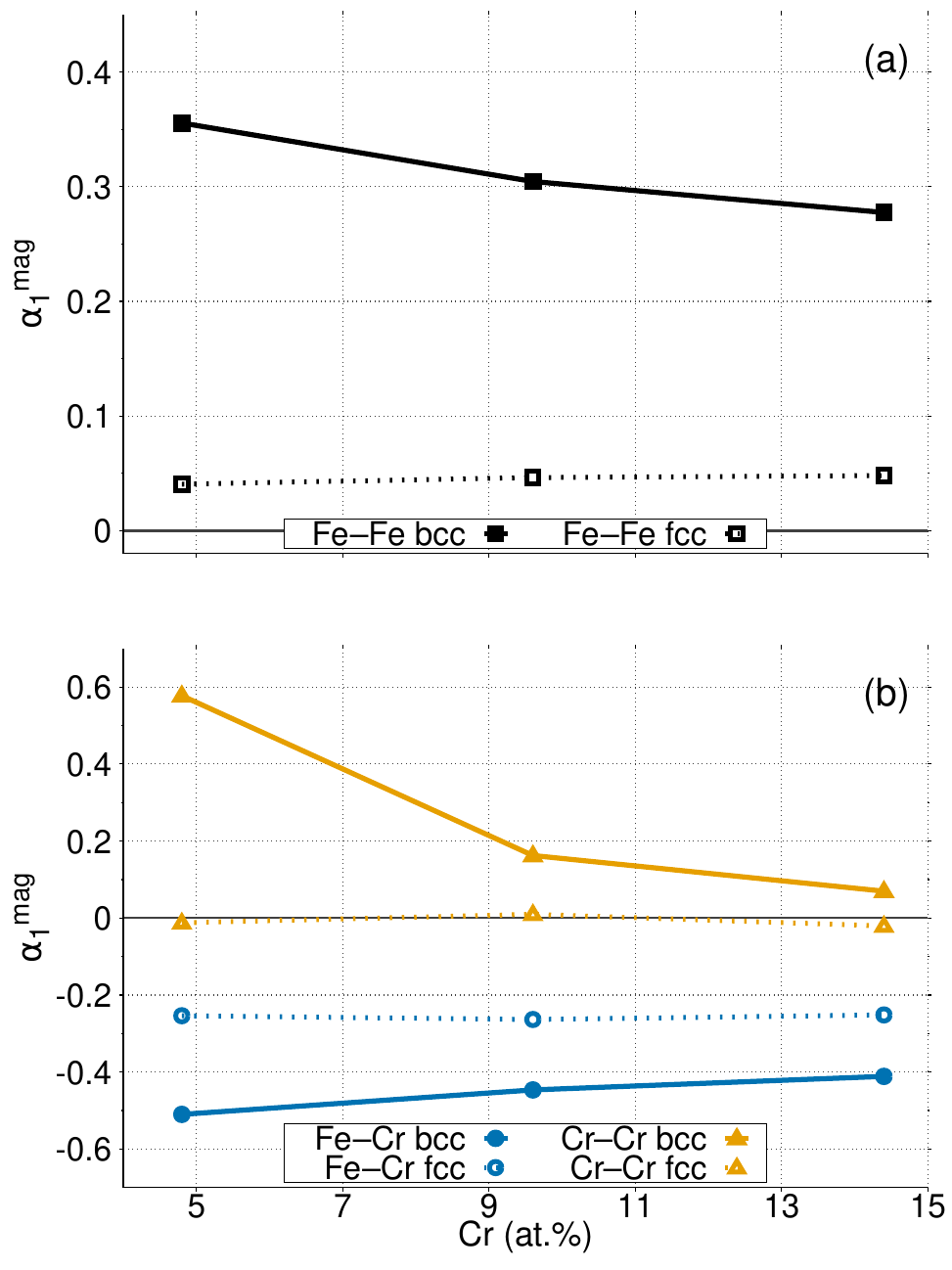}
    \caption{Magnetic pairwise short-range order ($\alpha_1^{mag}$) at approximate equilibrium volume at 1000 K as a function of Cr composition for all compositions and lattices, calculated using Eq.~\ref{sro_eq}. Panel (a) shows Fe-Fe $\alpha_1^{mag}$, while panel (b) shows Fe-Cr and Cr-Cr $\alpha_1^{mag}$.}
    \label{fig:SRO}
\end{figure}

For all temperatures and both lattice structures, the magnitude of the Fe and Cr moments decreases with increasing Cr concentration. This reduction is most pronounced for Cr in the bcc lattice at low temperatures. In this regime, the largest changes in nearest neighbor magnetic short-range order, $\alpha_1^{mag}$, between compositions are observed (see Fig.~\ref{fig:SRO}), indicating a strong correlation between $\alpha_1^{mag}$ variations and the moment reduction.  $\alpha_1^{mag}$ is defined as
\begin{equation}
\alpha_1^{mag} = \frac{1}{N}\sum_{i\neq j} \mathbf{e}_i \cdot \mathbf{e}_j
\label{sro_eq}
\end{equation}
where $\mathbf{e}_i$ and $\mathbf{e}_j$ represent the directions of the magnetic moments, expressed as unit vectors for lattice sites $i$ and $j$, respectively.

For Fe at low temperature, there is a significant difference between the moment magnitudes in the fcc and bcc lattices. This difference diminishes as temperature increases. The underlying reason is that the Fe moment in bcc decreases slightly with increasing temperature, whereas it increases in fcc. In both lattices, thermal expansion increases cell volume, which generally favors larger magnetic moments. In the bcc case, however, this effect is counteracted by a reduction in $\alpha_1^{mag}$ and the effect of vibrations of the lattice. The weakening of $\alpha_1^{mag}$ reduces the alignment in local magnetic environments, thereby diminishing the moment-enhancing effect of neighboring atoms. Vibrations can lead to instantaneous local distortions that shorten some Fe-Fe distances, thereby strongly suppressing the magnetic moments. This effect was already observed in collinear DLM-AIMD simulations in Ref. \cite{alling2016strong}, where it was shown that the electronic structure of bcc and fcc Fe becomes almost indistinguishable at the $\gamma-\delta$ transition temperatures.
For Cr, similar trends are observed as in the Fe case, with one notable exception: in the bcc lattice at high temperature, the Cr moments increase rather than decrease.

Fig.~\ref{fig:SRO} shows the magnetic short-range order, $\alpha_1^{mag}$, for Fe–Fe, Fe–Cr, and Cr–Cr nearest neighbor pairs in both the fcc and bcc lattices, evaluated near the equilibrium volume at 1000 K, where magnetic short-range order is expected to play a particularly important role \cite{ruban2012spin} in the stabilization of the bcc phase in this temperature range. 

Only the Fe magnetic moments were constrained during the ASD–AIMD simulations. Consequently, the Fe–Fe $\alpha_1^{mag}$ shown in Fig.~\ref{fig:SRO} arises directly from the exchange interactions included in the model. The Fe–Fe $\alpha_1^{mag}$ in the bcc lattice decreases with increasing Cr concentration, reflecting the dilution of Fe–Fe exchange interactions due to the Cr moments being treated as magnetic vacancies. In contrast, the Fe–Fe $\alpha_1^{mag}$ remains consistently low in the fcc simulations due to the intrinsically weaker exchange interactions in that lattice.

Since the Cr moments are not constrained and not included within the ASD framework, they do not explicitly contribute to magnetic ordering. They are instead initialized to 0 at each timestep of the AIMD simulation, and their direction and size evolve self-consistently, dependending on the constrained Fe-moment environment. The inclusion of Fe-Cr exchange, calculated using the method of Ref.~\cite{lindmaa2013exchange} and constraint of the Cr moments in Fe$_{85}$Cr$_{15}$ is evaluated in the Supplementary Material, where a neglible effect is found, showing the validity of treating Cr moments as secondary, environment-dependent moments dependent on the Fe magnetic configuration.

In fcc, the Fe–Cr and Cr–Cr pairs shown in Fig.~\ref{fig:SRO} exhibit only minor variations across the different compositions, which can be attributed to the low degree of magnetic ordering in fcc Fe~\cite{moruzzi1989magnetovolume,herper1999ab}. Antiferromagnetic ordering between Fe and Cr is found for all compositions, with $\alpha_1^{mag}$ values of approximately 0.25, whereas Cr-Cr $\alpha_1^{mag}$ is positive or null, reflecting how the Cr moments adapt to their local magnetic environment within the DFT calculations.

In the bcc structure, the $\alpha_1^{mag}$ tends to zero with increasing Cr composition, particularly for Cr–Cr pairs. The Cr–Cr $\alpha_1^{mag}$ decreases from slightly below 0.6 for Fe$_{95}$Cr$_{05}$ to slightly below 0.2 for Fe$_{90}$Cr$_{10}$, and further to approximately 0.1 for Fe$_{85}$Cr$_{15}$. Clear antiferromagnetic short-range order is observed for Fe–Cr pairs, while ferromagnetic short-range order is found for Fe–Fe pairs, with both decreasing as the Cr composition increases.

In the present treatment of the Cr moments, the results indicate that their magnetic behavior is strongly influenced by the surrounding Fe moments, as expected. This is clear from the bcc Fe-Cr and Cr-Cr $\alpha_1^{mag}$ results. At low Cr concentration, most nearest neighbors of a Cr atom are Fe atoms, so the antiferromagnetic Fe-Cr coupling determines the relaxed Cr orientation, leading to an antiferromagnetic Fe-Cr $\alpha_1^{mag}$ and a large ferromagnetic Cr-Cr. This interpretation is consistent with previous reports of antiferromagnetic Fe-Cr exchange interactions in dilute bcc Fe$_{1-x}$Cr$_x$~\cite{korzhavyi2009electronic}. The resulting ferromagnetic Cr-Cr pair correlations should therefore not be interpreted as evidence of a dominant ferromagnetic Cr-Cr interaction. Instead, they arise because neighboring Cr moments tend to adapt to a similar Fe-dominated magnetic environment. As the Cr concentration increases, the number of Cr-Cr neighbors rises relative to Fe-Cr and Fe-Fe neighbors, reducing the dominance of the Fe environment and leading to the observed decrease in Cr-Cr $\alpha_1^{mag}$, and a small increase in Fe-Cr $\alpha_1^{mag}$.

\subsection{Free energy and phase stability}

Detailed discussions of the step-by-step results obtained from simulations using DLM-AIMD, as well as SSTI, TTI, and VTI, that together lead to the full Gibbs free energies shown below, are provided in the Supplementary Materials.

\begin{figure}
    \centering
    \includegraphics[width=0.96\textwidth]{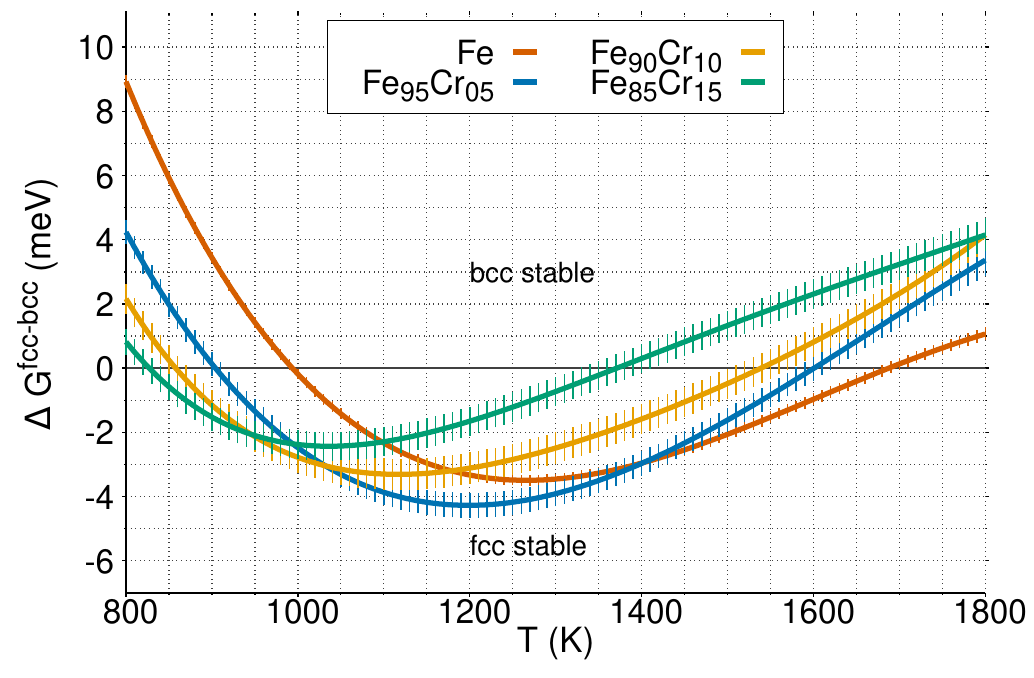}
    \caption{Gibbs free energy difference between fcc and bcc as a function of temperature. Fe, Fe$_{95}$Cr$_{05}$, Fe$_{90}$Cr$_{10}$, and Fe$_{85}$Cr$_{15}$ are shown in grey, black, blue, and red, respectively. The shaded areas indicate error bars corresponding to twice the standard deviation. Positive and negative y-values correspond to bcc and fcc stability, respectively. }
    \label{fig:dG_ASD}
\end{figure}

The Gibbs free energy difference between fcc and bcc at zero pressure, $\Delta G(x,T)=G^{fcc}(x,T)-G^{bcc}(x,T)$, for the compositions of pure Fe, Fe$_{95}$Cr$_{05}$, Fe$_{90}$Cr$_{10}$, and Fe$_{85}$Cr$_{15}$ is presented in Fig.~\ref{fig:dG_ASD}. Results for pure Fe are partially taken from Ref. \cite{Gambino2023Phase} and complemented with additional calculations to follow the same thermodynamic path used in this work.

The first realization is that for all studied compositions, the fcc $\gamma$-phase is only stable in a limited temperature range and with a maximum of a handful of meV/atom over the bcc $\alpha$- or $\delta$-phases. It demonstrates the unforgiving nature of the studied material system with respect to the need for completeness and accuracy in the used simulation methodology. 

If we start from pure Fe and inspect the curves with gradually larger Cr content, we find that adding Cr lowers the transition temperature at which the thermodynamic stability of the bcc and fcc phases change. This reduction originates from enhanced stabilization of the bcc phase relative to the fcc phase at higher temperatures which lowers the $\gamma-\delta$ transition. In contrast, at lower temperatures the opposite trend is observed, with the fcc phase beeing stabilized relative the bcc phase, again resulting in a lower transition temperature of the $\alpha-\gamma$ transition. This is indeed the same effect as experimentally observed as an anomaly of the shape of the $\gamma$-loop in the Fe-Cr phase diagram. The most significant change in the $\gamma-\delta$ transition temperature is found between  Fe$_{90}$Cr$_{10}$ and Fe$_{85}$Cr$_{15}$. In contrast, for the $\alpha-\gamma$ transition temperature, the most pronounced shift occurs between pure Fe and Fe$_{95}$Cr$_{05}$, with smaller decreases observed at higher Cr concentrations. The maximum stability of the fcc phase increases when moving from pure Fe, at -3.5 meV/atom  to -4.3 meV/atom at Fe$_{95}$Cr$_{05}$, but the trend is then reversed and decreases with further increases in Cr content.

\begin{figure}[]
    \centering
    \includegraphics[width=0.96\textwidth]{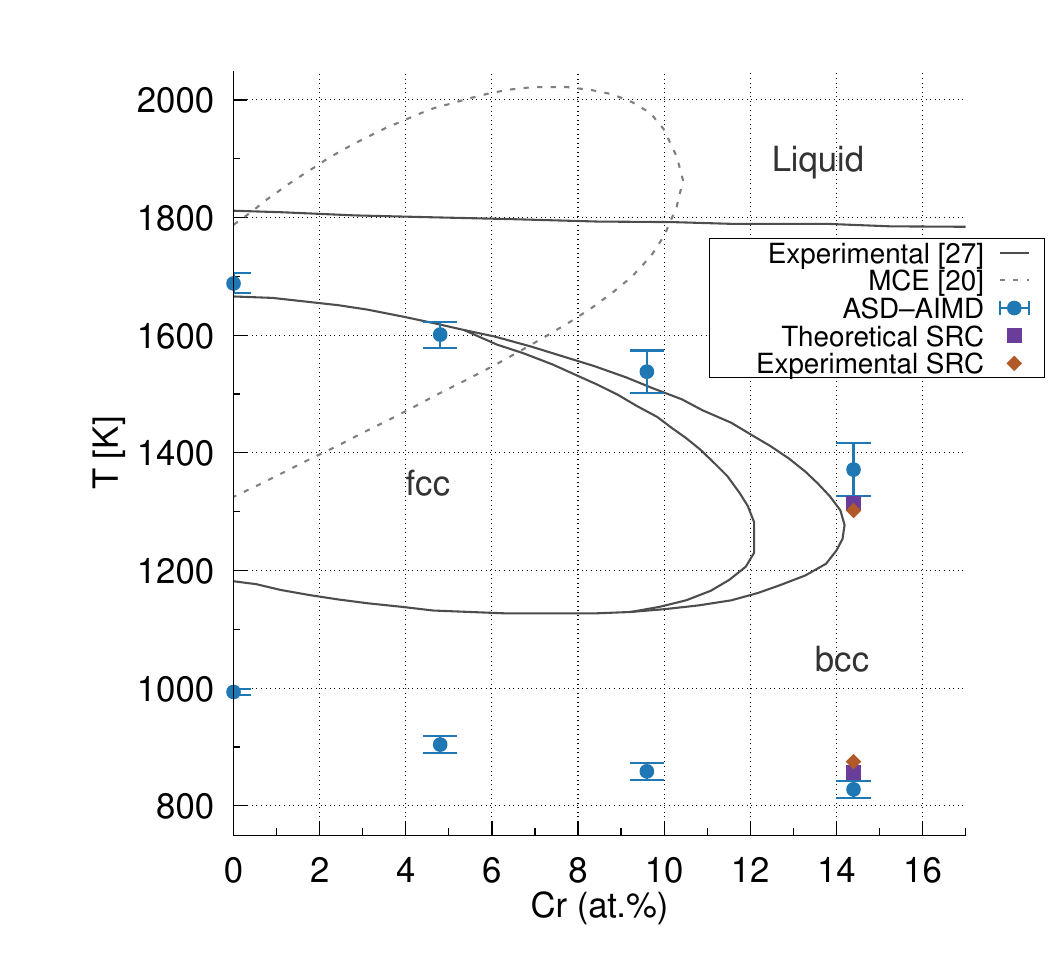}
    \caption{Phase diagram of the $\gamma$-loop region of the Fe-Cr system. Experimental data (solid lines) \cite{jacob2018revised} are compared to predicted fcc-bcc transitions in the present work using ASD-AIMD theoretical simulations (solid symbols). Error bars indicate the temperature uncertainty caused by two standard deviations of the statistical fluctuations of the simulations of the Gibbs free energy difference between fcc and bcc. Purple squares and red diamonds represent the transition temperatures when adding the calculated free energy from bcc short-range clustering (SRC) with theoretical \cite{korzhavyi2009electronic} and experimental \cite{mirebeau2019magnetic} SRC parameters respectively. Previous theoretical phase stability prediction \cite{lavrentiev2010magnetic} is presented as dashed lines}
    \label{fig:phase}
\end{figure}

When the calculated transition temperatures are plotted on the Fe–Cr phase diagram and compared with experimental data~\cite{jacob2018revised}, as shown in Fig.~\ref{fig:phase}, excellent agreement is obtained at the $\gamma-\delta$ transition at the top of the $\gamma$-loop. However, at lower transition temperatures, the calculations indicate an overstabilization of the fcc phase relative to experiments with systematic predictions of the $\alpha-\gamma$-transition around 200~K below the experimental observations. This deviation is observed across all compositions, suggesting that it originates from one or several of the approximations made in our ASD-AIMD simulations. 

Uncertainties in the fundamental exchange-correlation approximation used in our DFT calculations, GGA-PBE, can be part of the explanation, and so can the description of vibrational effects. However, as the deviations from experimental transition temperatures are very small at the $\gamma-\delta$ transition but larger at the  $\alpha-\gamma$ transition, the details of the description of magnetic short range order effects come into focus. In this aspect, both the finite size of our supercell, and the slight overestimation of the Curie temperature by our  bcc Fe–Fe magnetic exchange interactions~\cite{Yin.exchange.interactions, Gambino2023Phase}, give some uncertainties in the absolute value of the magnetic part of the free energies. 

At higher Cr concentrations, neglecting Fe–Cr and Cr–Cr interactions and representing chemical disorder using an SQS are expected to introduce larger uncertainties. For this reason, systems with Cr content above 15\% were not investigated. The potential errors associated with these approximations were instead assessed for the Fe$_{85}$Cr$_{15}$ composition. As discussed in the Methods section, the inclusion of Fe–Cr magnetic exchange interactions in our ASD-step of the simulations was found to have a negligible impact on the phase stability.

In 2010, M. Lavrentiev $et$ $al.$ \cite{lavrentiev2010magnetic} used magnetic cluster expansion to predict the phase stability of the gamma phase of Fe-Cr. These results are also included with dashed lines in Fig.~\ref{fig:phase}. In their approach, configurational disorder and magnetic fluctuations are treated explicitly, while the vibrational free energy is introduced separately via force-constant models derived from experimental elastic data. Due to this separate treatment, high-order cluster interactions are truncated, and anharmonic and spin-lattice couplings are not treated explicitly \cite{lavrentiev2010magnetic}. In contrast, the present work treats electronic structure, magnetic disorder and lattice dynamics simultaneously, thereby capturing magnetoelastic coupling and anharmonic effects intrinsically. It is apparent from Fig.~\ref{fig:phase} that this is very important in the Fe-Cr system and explains the qualitatively better results in our case, in particular the change in transition temperatures with Cr-content.

This far our analysis has assumed that the atomic configuration of both the bcc and the fcc alloy phases can be approximated as ideal random alloys represented by the SQS structures. This is a well-motivated starting point for analysis of the gamma-loop in Fe-Cr, as it involves relatively dilute alloy compositions and at relatively high temperatures. That is, conditions where configurational entropy can be expected to dominate the configurational degree of freedom and promote an atomic configuration that is close to random and well described by the SQS method. 

However, metastable bcc Fe$_{1-x}$Cr$_x$ alloys at Cr-richer compositions do exhibit an isostructural miscibility gap observed at temperatures below 1100 K if the equilibrium sigma-phase is avoided \cite{andersson1987thermodynamic}. This is supported by previous calculations observing positive mixing enthalpies in the bcc phase in both ordered and disordered magnetic states \cite{olsson2003ab, gambino2018lattice}. Outside the miscibility gap, SRC might lower the free energy of the bcc phase. In contrast, we are not aware of any observations or discussions about atomic SRC or ordering effects in fcc Fe-Cr phases. This motivates us as a first step to focus only on atomic SRC effects in the bcc phase while a detailed cluster-expansion study of all compositions for both structures must be left to a future work.

Among the most detailed theoretical works on atomic and magnetic thermodynamics in bcc Fe-Cr alloys are the work of Ruban $et$ $al.$  \cite{ruban2012first} and Korzhavyi $et$ $al.$ \cite{korzhavyi2009electronic}. They derived interatomic interactions in both the FM and DLM magnetic states as a function of composition. 

On the experimental side, a detailed study of SRC is done in Mirebeau $et$ $al.$ \cite{mirebeau2019magnetic} using mainly neutron diffuse scattering. They observed that Fe$_{85}$Cr$_{15}$ alloys annealed at T=1100 K and then quenched exhibit a mean atomic SRC on the first two shells of about $\alpha^{conf}_{12}= 0.025$ while alloys annealed at about T=700 K and then quenched has $\alpha^{conf}_{12}$ of about 0.065. 

As described in the Methods section, we perform atomic Monte-Carlo simulations of the configurational thermodynamics of the bcc Fe$_{0.85}$Cr$_{0.15}$ system using both directly the theoretical interactions from Ruban $et$ $al.$ \cite{ruban2012first} and Korzhavyi $et$ $al.$ \cite{korzhavyi2009electronic} and scaled interactions that reproduce experimental SRC parameters. Using thermodynamic integration from the high-temperature disordered limit downwards in temperature we obtain the free energy effect of SRC.

The resulting $\Delta G$ is shown in Fig.~S8 in the Supplementary Materials. The corresponding effects on the phase diagram are illustrated by additional data points at $x = 0.15$ in Fig.~\ref{fig:phase}.

It can be seen that the effect on the transition temperatures between bcc and fcc is relatively small, but still visible, at this composition. The difference between the two chosen sets of interatomic interactions used is very small. From the free energy curves, the magnitude of the effect is on the order of $1\,\mathrm{meV/atom}$.

In many systems, such an effect could reasonably be neglected. However, as previously discussed, the system studied here is highly sensitive to small energy differences. Test calculations using MC simulations at increased Cr content ($x = 0.20$) indicate, as expected, that the SRC effects increase by approximately $50\%$ compared to $x = 0.15$.

From this, we conclude that atomic SRC in the bcc phase contributes to closing the $\gamma$-loop, particularly causing the upward turn of its lower boundary at higher Cr content observed experimentally.

The results further indicate that using SQS to model the phases as random alloys is adequate in Fe$_{1-x}$Cr$_x$ up to $x = 0.15$. However, for higher Cr concentrations, this approach should be reconsidered in favor of a more comprehensive treatment, such as incorporating cluster expansion formalism alongside the present methods for including vibrational and magnetic effects for all considered structures.
 
\subsection{Discussion regarding the shape of the $\gamma$-loop}
 Before going into the discussion of the effect of Cr-content, let us recall that for pure Fe, the bcc-fcc-bcc transition before melting has been discussed in the following terms (see Ref.~\cite{Gambino2023Phase} and references therein): the ordered FM bcc state is lowest in total energy which decides the ground state of Fe. At increasing temperature, the loss of magnetic order reduces the energetic stabilization of bcc Fe, while magnetic entropy is suggested to stabilize fcc $\gamma-$Fe. The former effect is illustraded by the lower energy of the fcc structure as compared to the bcc structure in the magnetically maximally disordered state, here described with DLM calculations. It has also been suggested that the $\delta-$phase stability at yet higher temperatures is explained by vibrational contributions to the free energy of the more open bcc structure, particularly when the interatomic bonds of bcc are softened by magnetic disorder. Now let us discuss how these factors are influenced by Cr-addition.  
 
 The calculated $\gamma$-loop is shown in Fig.~\ref{fig:phase} to be in good agreement with the experimental-based reference data~\cite{jacob2018revised}. Indeed our simulations capture the anomalous difference as bcc is stabilized by Cr addition at the top while fcc is stabilized by Cr addition at the bottom of the loop. The top of the $\gamma-$loop, describing the $\gamma-\delta$ transition, is remarkably well reproduced in our calculations while we observe a systematic $\sim 200$ K underestimation of the $\alpha-\gamma$ transition temperature line at the bottom of the $\gamma-$loop. 

There are several possible sources for the underestimation of the transition temperatures at the bottom of the $\gamma-$loop. Our main message is that since the entire stability of the $\gamma-$phase arises from free energy differences of only a few meV/atom, also relatively small effects of different approximations can have a noticeable effect in the transition temperature. 

The finite cell size of 125 atoms do overstabilize magnetic order in the bcc structure at temperatures at the bottom of the $\gamma-$loop, as was previously shown by the prediction of too large Curie temperature for this cell size in ASD simulations \cite{Gambino2023Phase}. This will affect both the energy and the entropy of the simulations in opposite ways. However, we estimated finite-size effects on the free energy of the classical Heisenberg model to be below 2 meV/atom around the Curie temperature and favoring the bcc structure, suggesting that the small cell size does not explain the underestimation of the $\alpha-\gamma$ transition temperature. 

Other possible sources of uncertainties are the magnetic exchange interactions used for fcc \cite{Gambino2023Phase} and bcc \cite{Yin.exchange.interactions} in the ASD step of our simulations. Our implementations of the exchange interactions as distance dependent pair interactions are parametrizations of the real complex magnetic energetics, which give rise to some uncertainties even though they capture the main part of the magnetic energetics of the system. It should however be noted that it is the real DFT energies that enters the phase stability calculations while the parametrized interactions only enter indirectly through their guidance of the spin dynamics and thus which spin configurations that are probed by the DFT AIMD-step. 

The effects from the approximate treatment of exchange-correlation in DFT, and in particular the use of the PBE functional is another possible source of errors that could e.g. give a slight overstabilization of fcc over bcc crystal structures. However, it is not clear why this would give different errors at the top and bottom of the $\gamma-$loop. We have to leave a detailed investigation of these aspects, e.g. with simulations using other exchange-correlation functionals, to future works. 

We now turn to the unusual shape of the bottom of the $\gamma$-loop, where our calculations reproduce the downward slope of the $\alpha-\gamma$ boundary at low Cr contents, i.e. fcc stabilization by addition of a bcc-stabilizer (Cr). In principle, many contributions to the free energy differences could be the explanation to it; however, we start with  investigating if it can be explained by considering the system's energetics in the ordered and disordered magnetic states. 

\begin{figure}
    \centering
    \includegraphics[width=0.95\linewidth]{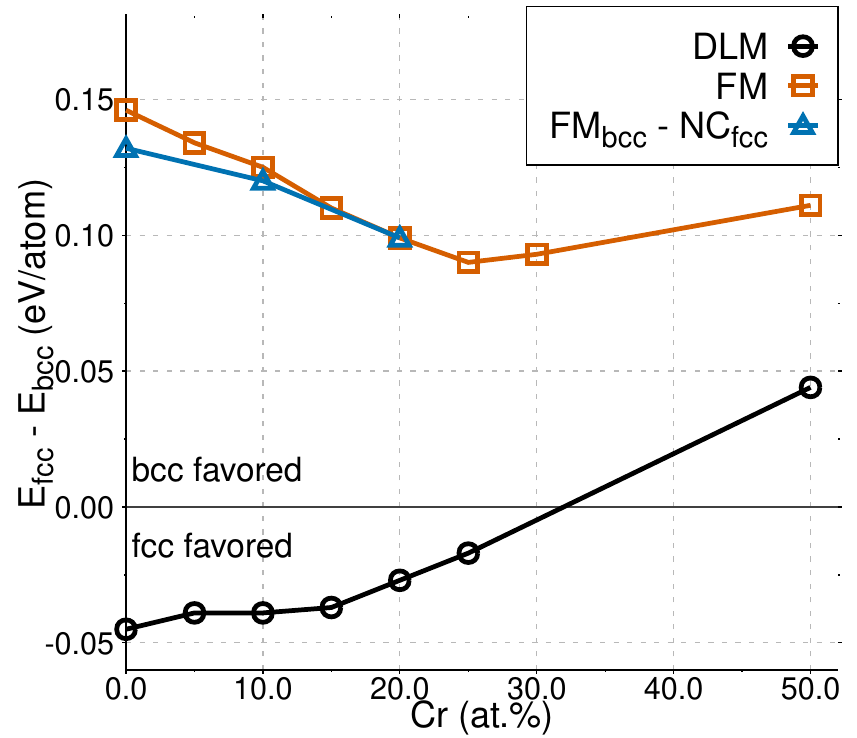}
    \caption{Energy differences between fcc and bcc for 0K FM and DLM simulations}
    \label{fig:dE}
\end{figure}

Fig.~\ref{fig:dE} presents the energy difference between the fcc and bcc phases using static DFT calculations at the volumes of our ASD-AIMD simulations at 1000 K. The simulations are done for three cases; first with both fcc and bcc being ferromagnetic; second as the energy difference between a magnetically ordered noncollinear fcc ground state (NC) and ferromagnetic bcc; third, when both phases are modeled using DLM as an approximation for the paramagnetic state. In the magnetically ordered cases, the bcc phase is strongly favored across compositions compared to both the ferromagnetic and NC fcc phases. Up to $x=0.20$, the NC fcc state has lower ground state energy compared to ferromagnetic fcc. The FM and NC converge to the approximatley the same value at 20\% Cr as the magnetic ground state search converge towards the FM order.

However, an important observation is that the bcc stability, although never beeing reversed, decrease with increasing Cr content from pure Fe up to $x=0.25$ after which the trend is reversed with again increasing bcc stability. 

In contrast, as mentioned above for the DLM calculations of pure Fe, the fcc phase is lower in energy. However, we find that increasing the Cr content decreases this energy preference of the fcc phase in the DLM magnetic state and at a Cr content of $x\sim 0.3$ bcc becomes energetically favourable also in DLM calculations.  

These two findings can indeed be seen as the main explanations of the different effects of Cr content on the top and bottom of the $\gamma-$loop. However, even if the DLM model is a good approximation of the magnetic state in the PM phase high above T$_C$ like the top of the $\gamma-$loop, the temperatures of the bottom of the $\gamma-$loop must be seen as disordered but with considerable magnetic short-range-order. Thus, in order to draw conclusions from Fig. 7 about the $\alpha-\gamma$ transition, one should consider a mixture of the DLM and the ordered magnetic results.  
 
Importantly, the reduction in bcc stability in the FM state is more pronounced than the corresponding reduction in fcc stability in the DLM state, implying a relative stabilization of the fcc phase at intermediate compositions. However, as the Cr content increases further, the destabilization of fcc in the DLM state becomes dominant, leading to a re-stabilization of the bcc phase.

Yang et al. \cite{yang2024insight} investigated the energetics of Fe-Cr at 0 K up to 12.5\% Cr and reported the same low-Cr trend. Since their study did not include the DLM state or higher Cr concentrations, however, the subsequent re-stabilization of bcc was not observed.

It should also not be forgotten, that at higher temperatures, particularly above the $\gamma$-loop, vibrational entropy becomes the primary stabilizing mechanism for the bcc phase, and effects of Cr-content on vibrations are not probed in these static calculation.

Finally we acknowledge that we do not attempt to resolve the small two-phase region between Fe-richer fcc and Cr-richer bcc that is observed in experiments with a maximum composition width of about 2 atomic percent. This does not influence any of our discussions above.  

\section{Conclusions }
In this work, we have investigated the temperature- and composition-dependent phase stability of Fe-Cr alloys in the Fe-rich part of the phase diagram using ASD-AIMD combined with thermodynamic integration. By treating magnetic disorder, lattice vibrations, and their coupling simultaneously, we reproduce the $\gamma$-loop of Fe-Cr and, in particular, the opposite effects of Cr addition on its lower and upper boundaries. The calculated $\gamma-\delta$ transition agrees well with experiment, whereas the $\alpha-\gamma$ transition is underestimated by about 200~K. Because the stability range of the fcc phase is governed by free-energy differences of only a few meV/atom, small uncertainties in the treatment of magnetic correlations, exchange interactions, finite-size effects, or exchange-correlation can substantially shift the calculated transition temperatures.  

The anomalous composition dependence of the $\gamma$-loop arises from a delicate balance in the energetics of different magnetic states for different Cr content. In magnetically ordered calculations, adding Cr initially reduces the preference for bcc, thereby stabilizing fcc. In the magnetically disordered state, the opposite trend occurs: the preference for fcc decreases with increasing Cr content and eventually shifts toward bcc. The competition between these two trends provide a microscopic explanation for why Cr addition initially expands the relative stability of fcc at the lower boundary of the $\gamma$-loop while stabilizing bcc at the upper boundary. 

The simulations provide insight into the local magnetic behavior of Fe-Cr at finite temperature. Finite Cr moments emerge spontaneously in both bcc and fcc simulations, despite not being explicitly constrained or induced in the ASD dynamics. Their magnitudes fluctuate more than those of the Fe moments and vary with composition and temperature. This behavior is consistent with Cr moments being comparatively sensitive to their local surroundings. The calculated magnetic short-range order varies with Cr concentration, as do the magnetic moment magnitudes and thermal expansion.

We have also assessed the effect of chemical SRC in the bcc phase. At 15 at.\% its free energy contribution is only on the order of 1 meV/atom, and the random-alloy SQS description therefore remains a reasonable approximation. Despite its small magnitude, this contribution reduces the stability range of the $\gamma$ phase. Its increasing importance with Cr content indicates that it contributes to the closure and upward turn of the lower bound of the $\gamma$-loop. A quantitative description further into the Fe-Cr phase diagram will therefore require magnetic, vibrational and configurational degrees of freedom to be treated on a comparable footing.

Overall, the results show that the anomalous shape of the Fe-Cr $\gamma$-loop is not controlled by a single free-energy contribution, but emerges from a competition among magnetic energetics, magnetic disorder and magnetic short-range correlations, lattice vibrations, and, at higher Cr concentrations, chemical short-range-clustering. Capturing the coupling between those degrees of freedom is therefore essential for a quantitative description through computer simulations of the phase stability in Fe-rich Fe-Cr alloys. The results of this work demonstrates that almost quantitative phase stability predictions for Fe-based alloys are coming within reach for first-principles based calculations. 

\section*{Acknowledgements}
Igor A. Abrikosov and Boburjon Mukhamedov are acknowledged for useful discussions. 
The computations were enabled by resources provided by National Academic Infrastructure for Supercomputing in Sweden (NAISS) partially funded by the Swedish Research Council through Grant Agreement No.  2022-06725. This work was supported by the Wallenberg Initiative Materials Science for Sustainability (WISE) funded by the Knut and Alice Wallenberg Foundation. 

DG acknowledges financial support from the Swedish Research Council (VR) through Grant No. 2023-00208 and from the ÅForsk Foundation through the Repatriation Grant 25-7.

B.A. acknowledges the financial support from the Swedish Research Council (VR) through Grant No.  2023-05194, from the Swedish Government Strategic Research Area in Materials Science on Functional Materials at Linköping University (Faculty Grant SFOMatLiU No. 2009-00971)

\section*{Generative AI Statement}
During the preparation of this manuscript, the authors used generative AI to assist with grammatical corrections, improving clarity, rewording, and shortening sections and paragraphs. All original drafts were written by the authors, who subsequently reviewed and verified all AI-assisted revisions. Generative AI was also used to assist in editing Gnuplot scripts for figure preparation. The authors take full responsibility for the content of the manuscript.
\newpage

\bibliographystyle{elsarticle-num}
\bibliography{references}

\end{document}